\documentclass{article}
\usepackage[utf8]{inputenc}
\usepackage{amsmath}
\usepackage{amsfonts}
\usepackage{amssymb}
\usepackage{hyperref}
\usepackage{braket}
\usepackage{authblk}
\usepackage[qm]{qcircuit2}
\usepackage{graphicx, caption}
\pdfmapfile{+sansmathaccent.map}

\usepackage[svgnames]{xcolor}

\let\vec\mathbf

\makeatletter
\renewcommand*\env@matrix[1][\arraystretch]{%
  \edef\arraystretch{#1}%
  \hskip -\arraycolsep
  \let\@ifnextchar\new@ifnextchar
  \array{*\c@MaxMatrixCols c}}
\makeatother

\title{Dementia Prediction Applying Variational Quantum Classifier}
\author[1]{Daniel Sierra-Sosa\textsuperscript{*}}
\author[2]{Juan Arcila-Moreno}
\author[3]{Begonya Garcia-Zapirain}
\author[3]{Cristian Castillo-Olea}
\author[1]{Adel Elmaghraby}
\affil[1]{Computer Science and Engineering Department, University of Louisville, KY, USA}
\affil[2]{Apolo Scientific Computing Center, Universidad EAFIT, Medell\'in, Colombia}
\affil[3]{eVida Research Group, University of Deusto, Bilbao, Spain}

\date{\today}                     
\begin{document}
\maketitle

\begin{abstract}
Dementia is the fifth cause of death worldwide with 10 million new cases every year. Healthcare applications using machine learning techniques have almost reached the physical limits while more data is becoming available resulting from the increasing rate of diagnosis. Recent research in Quantum Machine Learning (QML) techniques have found different approaches that may be useful to accelerate the training process of existing machine learning models and provide an alternative to learn more complex patterns. This work aims to report a real-world application of a Quantum Machine Learning Algorithm, in particular, we found that using the implemented version for Variational Quantum Classification (VQC) in IBM's framework Qiskit allows predicting dementia in elderly patients, this approach proves to provide more consistent results when compared with a classical Support Vector Machine (SVM) with a linear kernel using different number of features. 
\end{abstract}

\section{Introduction}
Dementia is a syndrome that comprises a range of forms that are indistinct, mixed and often co-exist, resulting in low rates of recognition by health-care professionals and patients \cite{WHO-ATLAS-mental}. The main repercussions of dementia are disability and dependency, creating physical, psychological, social and economic impacts on people who suffer it and their families. Worldwide 50 million people are diagnosed with Dementia, every year there are 10 million new cases of this disease, making it the fifth cause of death \cite{WHO-report-dementia}. This disease is recognized as a public health priority and entails a yearly cost of \$810 Billion USD, this implies a burden in the healthcare system.

There have been many efforts in creating predictive models to aid in the dementia diagnosis \cite{DMS_DataMining_multiple,Katako2018,Fisher2019}. The main premise is that statistical classification methods derived from data mining and machine learning could enhance computer assisted diagnosis. Some models like Support Vector Machines (SVM), Neural Networks and decision tree based models can provide improved accuracy, sensitivity and specificity of predictions evaluating data obtained by neuropsychological testing on dementia and other mental-health diseases \cite{DMS_BATTINENI2019100200}. Multiple advances in Machine Learning have proved that if healthcare models can scale to more realistic machine learning tasks, they will become an integral part of automated forecasting and decision-making systems.

The current state of classical healthcare applications using machine learning techniques have almost reached the physical limits for the solutions in terms of their speed, while the size of the available datasets is still increasing \cite{DMS_MOSCOSO2019101837}. Due to the massive data volume, conventional techniques will require vast resources to support healthcare professionals, consequently their scalability will be impaired by physical limitations in the future. Therefore, there is a need to create fast and feasible methods that strength healthcare professionals in the diagnostic process.

Quantum Technologies are growing at a fast pace, multiple fields have important developments achieved, implying that Quantum Computing is ready to reach the point where real world applications must be included into its scope. In particular, Quantum Machine Learning (QML) is one of the fields with further development \cite{Perdomo_Ortiz_2018} with several contributions including quantum inspired neural networks \cite{Arrazola_2019,Li2015InspiredNN,farhi2018classification}, hybridized low-depth Variational Quantum Circuits (VQC) \cite{ostaszewski2019quantum, Corcoles:nature:FS}, Nearest–Neighbour  \cite{Ruan2017}, Support Vector Machines (SVM) and algorithms for optimization and classification \cite{guerreschi2017practical, Distance-based:SchuldM, PhysRevLett.122.040504:MariaHilbert}. Indeed, these methods show that Quantum computers may be useful to accelerate the training process of existing learning models, also provide an alternative to learn more complex patterns in conventional datasets.

Some classical models have proven functionality in tasks related with mental-health disease prediction. Contributions including SVM  \cite{DSM_SVM,DSM_SVM_FS} and Neural Networks \cite{DMS_NeuralNetworks} have been presented, these models use demographic information, diagnose supported with cerebrospinal fluid biomarkers or imaging \cite{DMS_biomarks_FORLENZA2015455}, and neuropsychological tests including Mini Mental State Examination (MMSE) or Montreal Cognitive Assessment (MoCA) \cite{DMS_SORENSEN201866}. These applications could be enhanced by applying quantum computing techniques that are mathematically related with statistical classification methods, merging the individual advantages of these two technologies.

Provided the existing Near Intermediate Scale Quantum (NISQ) devices limitations, develop QML algorithms usually requires the usage of well-known or scaled-down datasets \cite{schuld2018supervised}, these are still far from being suitable for real applications. Thus, experiments that demonstrate the usage of health-care datasets with Quantum Classification models at a realistic scale are needed and valuable.

We present an application of Variational Quantum Classifier \cite{Corcoles:nature:FS} to a real dataset of Dementia patients comparing its prediction performance with a classical SVM classification model. This aims to provide useful information for future research in the application of real-world datasets to quantum machine learning techniques. This paper is arranged as follows: we present the preprocessing and feature selection techniques employed in our dataset, then we give a summarized description for VQC and SVM, following with the information related to the study case of dementia and we conclude with our results and discussions. 

\section{Materials and Methods}

Quantum Machine Learning techniques are currently impaired by NISQ devices limitations \cite{Preskill2018quantumcomputingin}. Thus, many proposed QML applications rely on using well-known datasets, where pre-processing techniques are standard \cite{Distance-based:SchuldM,schuld2018circuitcentric,kerenidisMNIST2018quantum}. 

\subsection{Pre-processing and Feature Selection}

The first step is to ensure that any input in the model are in the same scale, this will provide the same weight to all features involved in the classification process. This step is required in most of the machine learning techniques, constituting a preprocessing pipeline  \cite{Hands-On_MLSKTF}. For Quantum Machine Learning applications, pre-processing data has a particular relevance provided the system constraints. Preprocessing data is a fundamental step to represented classical data into quantum states. In this particular proposal we used min-max normalization.

Besides normalizing data, we conducted feature selection techniques, removing less relevant or redundant information, reducing data dimensionality  \cite{LIU1998333}. In medical datasets, these techniques are common given that usually medical features induce noise in the models. Algorithms like Principal Component Analysis (PCA) are widely used for this purpose, including QML applications and processing \cite{Lloyd_2014,QuantumCompressionPCA:2019}. We based our feature selection methodology in a variable ranking, calculating a score obtained when applying Gradient Boosting \cite{Xu:2014:GBF:2623330.2623635}, Random forest and Extra Trees that minimizes overfitting the data, and K-Best. We obtained a subset of features from our dataset that based on the top features of the calculated score.

\subsection{VQC}

Variation Quantum Classfier (VQC) is an algorithm that allows to obtain experimental results in NISQ devices without the need to perform additional error-correction techniques. This method is a hybrid approach where the parameters are optimized and updated in a classical computer, making the optimization process without increasing the coherence times needed. The cost function calculation is base on iterative measurements from the device serve as error mitigation, by including noisy measurements into optimization calculations \cite{Corcoles:nature:FS}. Schuld et.al. \cite{PhysRevLett.122.040504:MariaHilbert,schuld2018circuitcentric} showed that mapping features to quantum states using amplitude encoding, is a suitable option to pre-process data when using VQC, provided that data is low dimensional or its structure allows for efficient approximate preparation. 

One of the key components from this method is the feature map definition, which maps data into a potentially vastly higher-dimensional Hilbert space of a quantum system \cite{PhysRevLett.122.040504:MariaHilbert} allowing to perform efficient computations over non-linear basic functions on a possibly intractably large space, the feature space. 

\begin{figure}
    \centering
    \includegraphics[width=0.7\textwidth]{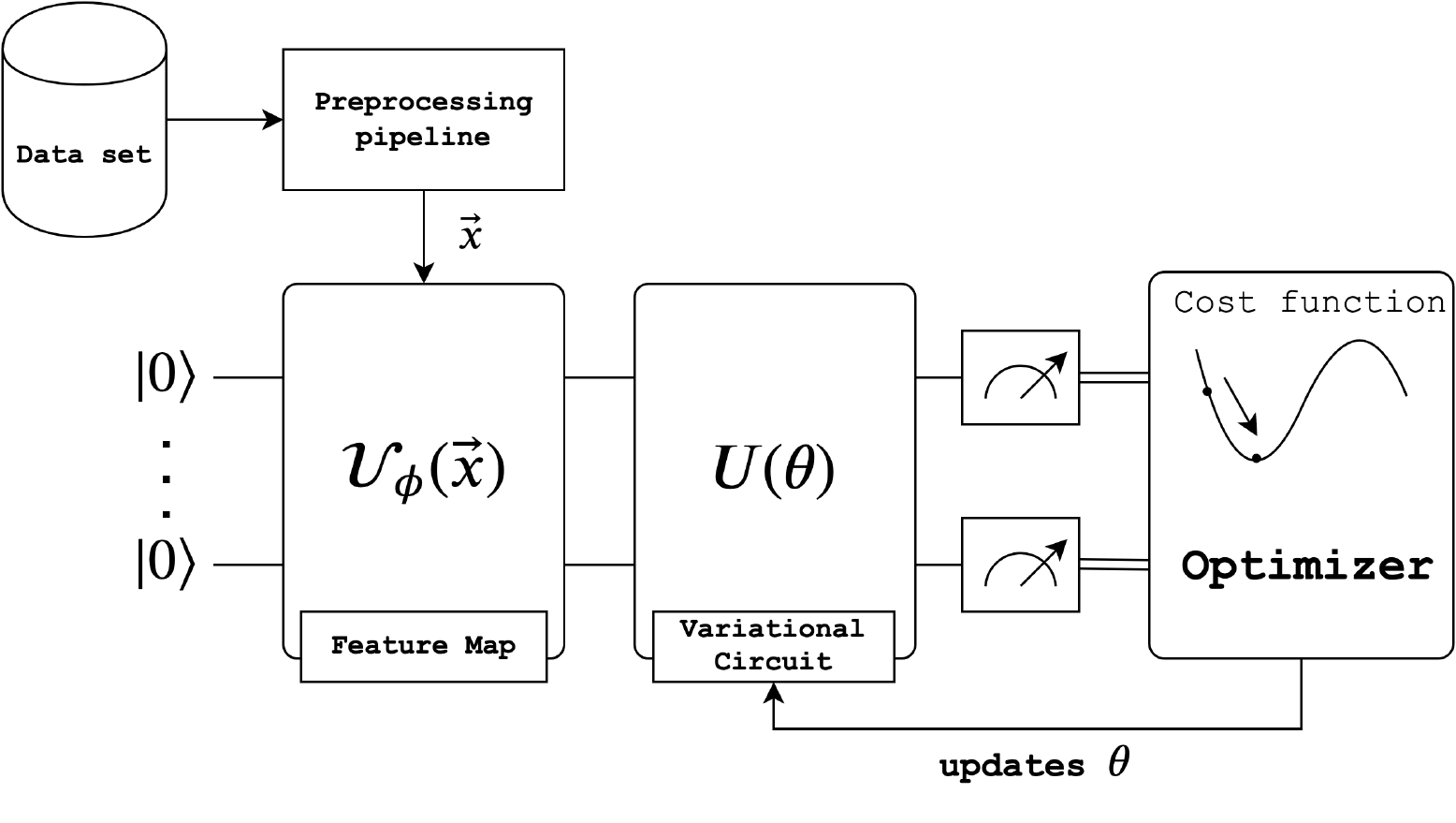}
    \caption{Schematic View of VQC Algorithm}
    \label{fig:vqc}
\end{figure}

Havlíček, et.al. \cite{Corcoles:nature:FS} proposed a VQC method with two main elements presented in Figure \ref{fig:vqc}. First, a feature map that works as a fixed black-box encoding classical data $\vec{x_i}$ into a quantum states $\ket{\Psi(\vec{x_i})}$, by applying transformations to the ground state $\ket{0}^n$ using products of single and two-qubits unitaries phase-gates. In specific, the experimental implementation of the authors result in a unitary gate ${\cal U}_\phi(\vec{x})=U_{\phi(\vec{x})}H^{\otimes n}U_{\phi(\vec{x})}H^{\otimes n}$ where $H$ represent the Hadamard gate and 
$$U_{\phi(\vec{x})}=exp\left(i\sum_{S\subseteq[n]}\phi s(\vec{x})\prod_{i\in S}Z_i \right),$$
 is a diagonal gate in the Pauli-Z basis. Second, a short depth unitary $U(\theta)$ circuit with $l$ layers of $\theta$-parameters, optimized during training by minimizing a cost function in a classical device, and tuning  $\theta$ iteratively. Using parameterized quantum circuits known as Quantum Circuit Learning (QCL), implies the usage of an exponential number of functions with respect to the number of qubits from the parameterized circuit, this is intractable on classical computers, therefore, allows to represent more complex functions than the classical counterparts \cite{10Spin:PhysRevA.98.032309}. 
 
\subsection{SVM}

Support Vector Machine (SVM) is a Machine Learning method capable of performing linear and nonlinear classification over small or medium sized datasets with a high level of complexity in a suitable amount of time \cite{Hands-On_MLSKTF_SVM}. The fundamental idea behind this method is to perform an optimization task finding an optimal hyperplane which classifies new examples and data points relying on a classification margin by using instances located at its edge called Support vectors. This methodology allows to control the balance of the decision boundary in the hyperplane, preventing margin violations when classifying new samples, and acting as a regularization type. Formally, finding a classification margin imply solving a convex quadratic problem with linear constraints, also known as Quadratic Programming, that follows the general formulation described in Equation \eqref{eq:svm1} 

\begin{equation} \label{eq:svm1}
\begin{aligned}
& \underset{p}{\text{Minimize}}
& & \frac{1}{2}p^{T} \cdot H \cdot p + f^{T} \cdot p  \\
& \text{subject to}
& & A \cdot p \leq b
\end{aligned}
\end{equation}

\begin{equation}\notag
where
\begin{cases}
\boldsymbol{p} & \text{is an  $n_p$-dimensional vector ($n_p$ = number of parameters),}\\
\boldsymbol{H} & \text{is an  $n_p \times n_p$  matrix,}\\
\boldsymbol{f} & \text{is an  $n_p$-dimensional vector,}\\
\boldsymbol{A} & \text{is an  $n_c \times n_p$ matrix ($n_c$ = number of constraints),}\\
\boldsymbol{b} & \text{is an  $n_c$-dimensional vector.}\\
\end{cases}
\end{equation}    

\begin{figure}
    \centering
    \includegraphics[width=0.5\textwidth]{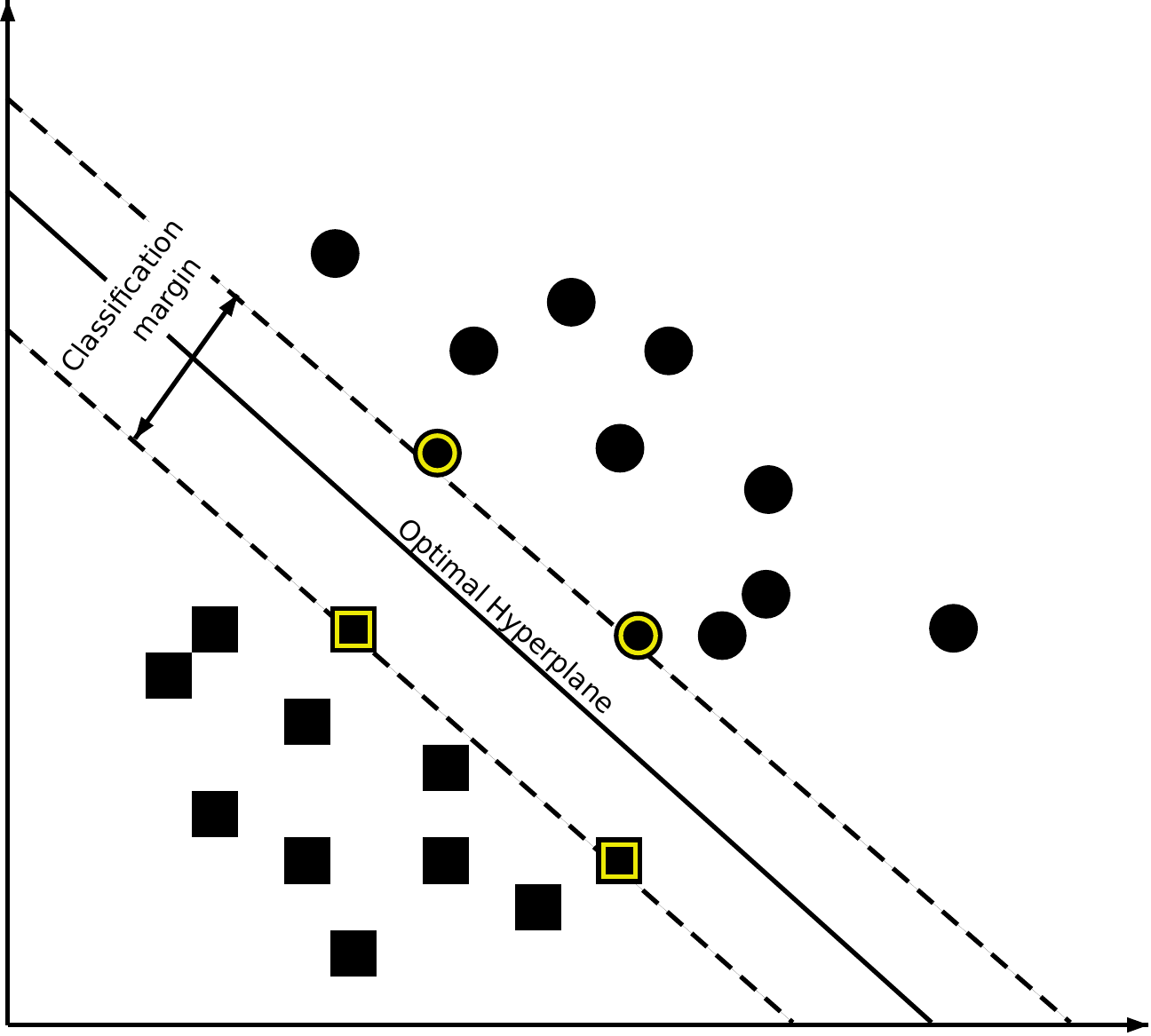}
    \caption{Elements of SVM machine}
    \label{fig:svm}
\end{figure}

In the Figure \ref{fig:svm} we can observe the different elements of SVM classifying a linear separable dataset using two dimensions. The remarked points are the instances for the support vectors. An important advantage of SVM method is that it makes possible to transform data into a higher dimensional space where it can be easily separable without performing the actual calculations, leaving the mapping completely implicit. This is known as “\textit{Kernel Trick}” and is possible following \textit{Mercer’s theorem}: It exist a function $\phi$ that maps the points $a$ and $b$ into another possibility beastly dimensional space such that a kernel function $K$ that complaints certain conditions, can be expressed as $K(a,b) = \phi(a)^{T} \cdot \phi(b)$, allowing to employ the Kernel $K$ instead of completely compute the dot product. Depending on the selected kernel, this can get the same result as if you added polynomial features to classify the data \cite{Hands-On_MLSKTF_SVM,SVMFeaturePoly-2007}.

\section{Case Study: Dementia prediction on elder patients}
A dementia study was conducted in elder adults at the General Hospital of Tijuana. In 2017 the total elderly population between 65 and 90 years of age in Tijuana was 85,529. Out of these patients, 65\% were treated in the General Hospital of Tijuana. This Hospital is a public institution for a population with limited resources. The hospital serves people in Baja California region specifically Tijuana, Mexicali, Tecate, Rosarito, and Ensenada. Table \ref{tab:dataset} presents the demographic data.

\begin{table}[]
\centering
\begin{tabular}{|c|l|}
\hline
\textbf{Gender (\%)} & $76\%$ (Female) \\ \hline
\textbf{Average Schooling (Years)} & $3.46 \pm 3.8767$ \\ \hline
\textbf{Average Age (Years)} & $78.90 \pm 7.7939$ \\ \hline
\textbf{Avg. Number of Chronic Diseases} & $2.96 \pm 1.3242$ \\ \hline
\end{tabular}
\caption{Demographic information of employed Dementia dataset from the General Hospital of Tijuana }
\label{tab:dataset}
\end{table}

The data collection for the database creation contains information of 166 patients with 99 variables of each.   Patients with physical or psychological dependence were excluded from this study. Collected data includes clinical history, psychological tests; comorbidity (Charlson), functional ability (Barthel and Lawton), malnutrition (MNA validated test), pharmacology, biochemical data, and demographic data.

\section{Results}

Data classification was performed by using the implemented version of VQC in IBM’s framework \textit{Qiskit} version 0.11.1 and executed in the provided simulator Aer version 0.2.3 \cite{Qiskit}. Every combination of the experiments were executed 1024 shots, using the implemented version of the Cobyla optimizer  \cite{Pow94:proc}  through the same framework. We conducted tests with different number of qubits ranging from 2 to 5, in each case we compared the accuracy, precision, recall and F1-Score.

We found that using Variational Quantum Classification method performs well predicting the presence of Dementia in patients compared to classical classification models, in specific a SVM with linear kernel. Our experiments show that Quantum Machine Learning are suitable to classify real-world data achieving close enough results compared to these classical approaches. 

\begin{figure}
    \centering
    \includegraphics[width=\textwidth]{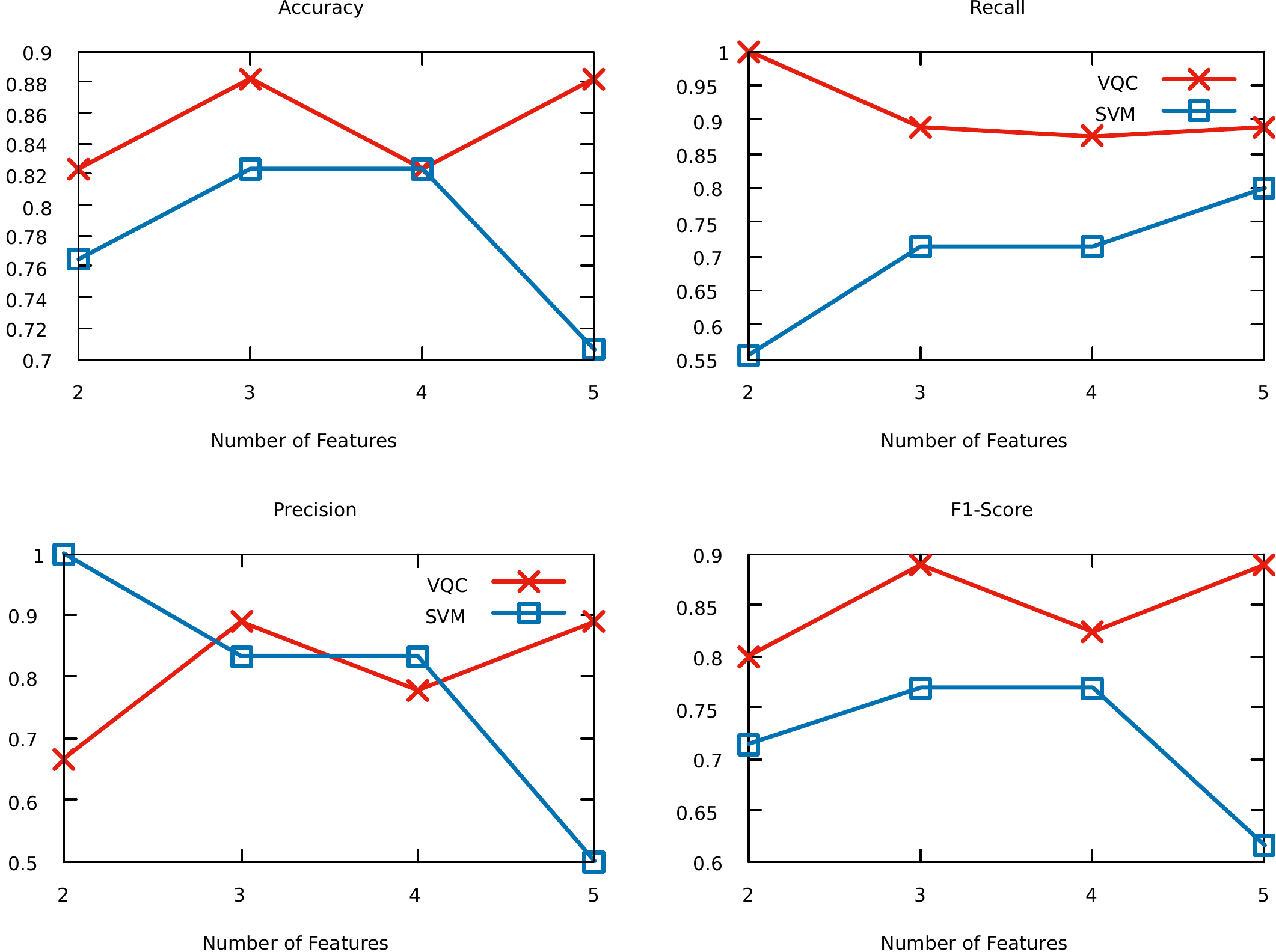}
    \caption{Dementia prediction metrics for different number of Features}
    \label{fig:results}
\end{figure}

Given the limitations on current NISQ devices, we selected a reduced number of features to predict dementia, taking into account the actual number of qubits and algorithm execution time. The results from these experiments are summarized in Figure \ref{fig:results}, where it can be seen that VQC in general outperform SVM and the results are more consistent when changing the number of features. Therefore, with the evaluated metrics we found that using Variational Quantum Classification method is a suitable approach to predict the presence of Dementia. Moreover, our experiments indicate that Quantum Machine Learning techniques could be used to classify real-world data archiving close enough results compared with classical approaches. 

\section{Conclusions}

Dementia is a major concern in public health, therefore several research efforts have been conducted including topics that are addressed using statistics, data mining and machine learning techniques. Every year more data is becoming available from the increased diagnosis rate. The data availability has lead to the emergence from Machine learning methods, which are nowadays an extremely valuable tool for healthcare professionals to understand and mitigate this and other conditions. 

In particular, major concern diseases research is enhancing due to this data increment, the next stage will continue to explore new machine learning techniques that performs equally or better than current methods, optimizing the computational resources when obtaining predictions. Quantum Machine Learning Algorithms have proven theoretical advantages in terms of computational complexity when compared with some classical counterparts. A milestone to pursue is to achieve quantum advantages in real applications, our results suggest that dementia prediction for health-mental related diseases is a feasible and valuable example, we foresee that in the future QML will enhance medical data assessment, providing more efficient support to healthcare professionals.

This work reports an application of a Quantum Machine Learning Algorithm, using the capabilities of the methods available through the current devices and frameworks for quantum computing. In particular, we materialize an application of VQC for a real-world dataset on dementia patients. We compared VQC performance with classical SVM classification model using linear kernel. Our results showed that VQC performs consistently better when using different number of features. Although some SVM cases show better precision classifying specific classes, this method is not as proficient when using a limited number of features. In general, VQC results demonstrate that it is a promising technique when quantum devices grown its capabilities, attending the future necessities of the healthcare system. Even if according to our experiments, current state-of-art techniques can not be yet replaced with VQC for dementia prediction, we foresee that these methods will be of common use in the near future. We are currently working in exploring new approaches that could enhance the outcomes of these quantum techniques, evaluating the execution advantages of applying algorithms in different environments. 

\section*{Data Availability}
The data that support the findings of this study are available from the corresponding author upon justified request by email.

\section*{References}
\bibliography{main.bib} 

\begin{thebibliography}{10}

\bibitem{WHO-ATLAS-mental}
{World Health Organization}.
\newblock {\em Mental Health ATLAS 2017}.
\newblock World Health Organization, 2018.

\bibitem{WHO-report-dementia}
{World Health Organization and Alzheimer’s Disease International}.
\newblock {\em Dementia: a public health priority}.
\newblock World Health Organization, 2012.

\bibitem{DMS_DataMining_multiple}
João Maroco, Dina Silva, Ana Pina~Rodrigues, Manuela Guerreiro, Isabel
  Santana, and Alexandre Mendonça.
\newblock Data mining methods in the prediction of dementia: A real-data
  comparison of the accuracy, sensitivity and specificity of linear
  discriminant analysis, logistic regression, neural networks, support vector
  machines, classification trees and random forests.
\newblock {\em BMC research notes}, 4:299, 08 2011.

\bibitem{Katako2018}
Audrey Katako, Paul Shelton, Andrew~L. Goertzen, Daniel Levin, Bohdan Bybel,
  Maram Aljuaid, Hyun~Jin Yoon, Do~Young Kang, Seok~Min Kim, Chong~Sik Lee, and
  Ji~Hyun Ko.
\newblock Machine learning identified an alzheimer's disease-related fdg-pet
  pattern which is also expressed in lewy body dementia and parkinson's disease
  dementia.
\newblock {\em Scientific Reports}, 8(1):13236, 2018.

\bibitem{Fisher2019}
Charles~K. Fisher, Aaron~M. Smith, Jonathan~R. Walsh, Adam~J. Simon, Chris
  Edgar, Clifford~R. Jack, and David Holtzman.
\newblock Machine learning for comprehensive forecasting of alzheimer's disease
  progression.
\newblock {\em Scientific Reports}, 9(1):13622, 2019.

\bibitem{DMS_BATTINENI2019100200}
Gopi Battineni, Nalini Chintalapudi, and Francesco Amenta.
\newblock Machine learning in medicine: Performance calculation of dementia
  prediction by support vector machines (svm).
\newblock {\em Informatics in Medicine Unlocked}, 16:100200, 2019.

\bibitem{DMS_MOSCOSO2019101837}
Alexis Moscoso, Jesús Silva-Rodríguez, Jose~Manuel Aldrey, Julia Cortés,
  Anxo Fernández-Ferreiro, Noemí Gómez-Lado, Álvaro Ruibal, and Pablo
  Aguiar.
\newblock Prediction of alzheimer's disease dementia with mri beyond the
  short-term: Implications for the design of predictive models.
\newblock {\em NeuroImage: Clinical}, 23:101837, 2019.

\bibitem{Perdomo_Ortiz_2018}
Alejandro Perdomo-Ortiz, Marcello Benedetti, John Realpe-G{\'{o}}mez, and Rupak
  Biswas.
\newblock Opportunities and challenges for quantum-assisted machine learning in
  near-term quantum computers.
\newblock {\em Quantum Science and Technology}, 3(3):030502, jun 2018.

\bibitem{Arrazola_2019}
Juan~Miguel Arrazola, Thomas~R Bromley, Josh Izaac, Casey~R Myers, Kamil
  Brádler, and Nathan Killoran.
\newblock Machine learning method for state preparation and gate synthesis on
  photonic quantum computers.
\newblock {\em Quantum Science and Technology}, 4(2):024004, Jan 2019.

\bibitem{Li2015InspiredNN}
Jianping Li.
\newblock Quantum-inspired neural networks with application.
\newblock {\em Open Journal of Applied Sciences}, 05:233--239, 01 2015.

\bibitem{farhi2018classification}
Edward Farhi and Hartmut Neven.
\newblock Classification with quantum neural networks on near term processors,
  2018.

\bibitem{ostaszewski2019quantum}
Mateusz Ostaszewski, Edward Grant, and Marcello Benedetti.
\newblock Quantum circuit structure learning, 2019.

\bibitem{Corcoles:nature:FS}
Vojtěch Havlíček, A.~Córcoles, Kristan Temme, Aram Harrow, Abhinav Kandala,
  Jerry Chow, and Jay Gambetta.
\newblock Supervised learning with quantum-enhanced feature spaces.
\newblock {\em Nature}, 567:209--212, 03 2019.

\bibitem{Ruan2017}
Yue Ruan, Xiling Xue, Heng Liu, Jianing Tan, and Xi~Li.
\newblock Quantum algorithm for k-nearest neighbors classification based on the
  metric of hamming distance.
\newblock {\em International Journal of Theoretical Physics},
  56(11):3496--3507, Nov 2017.

\bibitem{guerreschi2017practical}
Gian~Giacomo Guerreschi and Mikhail Smelyanskiy.
\newblock Practical optimization for hybrid quantum-classical algorithms, 2017.

\bibitem{Distance-based:SchuldM}
M.~Schuld, M.~Fingerhuth, and Francesco Petruccione.
\newblock Implementing a distance-based classifier with a quantum interference
  circuit.
\newblock {\em EPL (Europhysics Letters)}, 119:60002, 09 2017.

\bibitem{PhysRevLett.122.040504:MariaHilbert}
Maria Schuld and Nathan Killoran.
\newblock Quantum machine learning in feature hilbert spaces.
\newblock {\em Phys. Rev. Lett.}, 122:040504, Feb 2019.

\bibitem{DSM_SVM}
I.~{\'A}lvarez, J.~M. G{\'o}rriz, J.~Ram{\'i}rez, D.~Salas-Gonzalez,
  M.~L{\'o}pez, F.~Segovia, C.~G. Puntonet, and B.~Prieto.
\newblock Alzheimer's diagnosis using eigenbrains and support vector machines.
\newblock In Joan Cabestany, Francisco Sandoval, Alberto Prieto, and Juan~M.
  Corchado, editors, {\em Bio-Inspired Systems: Computational and Ambient
  Intelligence}, pages 973--980, Berlin, Heidelberg, 2009. Springer Berlin
  Heidelberg.

\bibitem{DSM_SVM_FS}
Esther Bron, Marion Smits, John van Swieten, Wiro Niessen, and Stefan Klein.
\newblock Feature selection based on svm significance maps for classification
  of dementia.
\newblock In Guorong Wu, Daoqiang Zhang, and Luping Zhou, editors, {\em Machine
  Learning in Medical Imaging}, pages 272--279, Cham, 2014. Springer
  International Publishing.

\bibitem{DMS_NeuralNetworks}
Sylvester~Olubolu Orimaye, Jojo Sze-Meng Wong, and Chee~Piau Wong.
\newblock Deep language space neural network for classifying mild cognitive
  impairment and alzheimer-type dementia.
\newblock {\em PLOS ONE}, 13(11):1--15, 11 2018.

\bibitem{DMS_biomarks_FORLENZA2015455}
Orestes~V. Forlenza, Marcia Radanovic, Leda~L. Talib, Ivan Aprahamian, Breno~S.
  Diniz, Henrik Zetterberg, and Wagner~F. Gattaz.
\newblock Cerebrospinal fluid biomarkers in alzheimer's disease: Diagnostic
  accuracy and prediction of dementia.
\newblock {\em Alzheimer's \& Dementia: Diagnosis, Assessment \& Disease
  Monitoring}, 1(4):455 -- 463, 2015.

\bibitem{DMS_SORENSEN201866}
Lauge Sørensen and Mads Nielsen.
\newblock Ensemble support vector machine classification of dementia using
  structural mri and mini-mental state examination.
\newblock {\em Journal of Neuroscience Methods}, 302:66 -- 74, 2018.
\newblock A machine learning neuroimaging challenge for automated diagnosis of
  Alzheimer’s disease.

\bibitem{schuld2018supervised}
Maria Schuld and Francesco Petruccione.
\newblock {\em Supervised Learning with Quantum Computers}, volume~17.
\newblock Springer, 2018.

\bibitem{Preskill2018quantumcomputingin}
John Preskill.
\newblock Quantum {C}omputing in the {NISQ} era and beyond.
\newblock {\em {Quantum}}, 2:79, August 2018.

\bibitem{schuld2018circuitcentric}
Maria Schuld, Alex Bocharov, Krysta Svore, and Nathan Wiebe.
\newblock Circuit-centric quantum classifiers, 2018.

\bibitem{kerenidisMNIST2018quantum}
Iordanis Kerenidis and Alessandro Luongo.
\newblock Quantum classification of the mnist dataset via slow feature
  analysis, 2018.

\bibitem{Hands-On_MLSKTF}
Aurlien Gron.
\newblock {\em Hands-On Machine Learning with Scikit-Learn and TensorFlow:
  Concepts, Tools, and Techniques to Build Intelligent Systems}, chapter~2,
  page~65.
\newblock O'Reilly Media, Inc., 1st edition, 2017.

\bibitem{LIU1998333}
Huan Liu and Rudy Setiono.
\newblock Some issues on scalable feature selection1this is an extended version
  of the paper presented at the fourth world congress of expert systems:
  Application of advanced information technologies held in mexico city in march
  1998.1.
\newblock {\em Expert Systems with Applications}, 15(3):333 -- 339, 1998.

\bibitem{Lloyd_2014}
Seth Lloyd, Masoud Mohseni, and Patrick Rebentrost.
\newblock Quantum principal component analysis.
\newblock {\em Nature Physics}, 10(9):631–633, Jul 2014.

\bibitem{QuantumCompressionPCA:2019}
Chao-Hua Yu, Fei Gao, Song Lin, and Jb~Wang.
\newblock Quantum data compression by principal component analysis.
\newblock {\em Quantum Information Processing}, 18, 08 2019.

\bibitem{Xu:2014:GBF:2623330.2623635}
Zhixiang Xu, Gao Huang, Kilian~Q. Weinberger, and Alice~X. Zheng.
\newblock Gradient boosted feature selection.
\newblock In {\em Proceedings of the 20th ACM SIGKDD International Conference
  on Knowledge Discovery and Data Mining}, KDD '14, pages 522--531, New York,
  NY, USA, 2014. ACM.

\bibitem{10Spin:PhysRevA.98.032309}
K.~Mitarai, M.~Negoro, M.~Kitagawa, and K.~Fujii.
\newblock Quantum circuit learning.
\newblock {\em Phys. Rev. A}, 98:032309, Sep 2018.

\bibitem{Hands-On_MLSKTF_SVM}
Aurlien Gron.
\newblock {\em Hands-On Machine Learning with Scikit-Learn and TensorFlow:
  Concepts, Tools, and Techniques to Build Intelligent Systems}, chapter~5,
  pages 161--163.
\newblock O'Reilly Media, Inc., 1st edition, 2017.

\bibitem{SVMFeaturePoly-2007}
Iv\'an Mej\'ia-Guevara, \'Angel Kuri-Morales, Domingo Mery, and Josef Kittler.
\newblock Mp-polynomial kernel for training support vector machines.
\newblock In {\em Progress in Pattern Recognition, Image Analysis and
  Applications}, pages 584--593, Berlin, Heidelberg, 2007. Springer Berlin
  Heidelberg.

\bibitem{Qiskit}
H{\'e}ctor Abraham, Ismail~Yunus Akhalwaya, Gadi Aleksandrowicz, Thomas
  Alexander, Gadi Alexandrowics, Eli Arbel, Abraham Asfaw, Carlos Azaustre,
  Panagiotis Barkoutsos, George Barron, Luciano Bello, Yael Ben-Haim, Lev~S.
  Bishop, Samuel Bosch, David Bucher, CZ, Fran Cabrera, Padraic Calpin, Lauren
  Capelluto, Jorge Carballo, Chun-Fu Chen, Adrian Chen, Richard Chen, Jerry~M.
  Chow, Christian Claus, Andrew~W. Cross, Abigail~J. Cross, Juan Cruz-Benito,
  Cryoris, Chris Culver, Antonio~D. C{\'o}rcoles-Gonzales, Sean Dague, Matthieu
  Dartiailh, Abd{\'o}n~Rodr{\'\i}guez Davila, Delton Ding, Eugene Dumitrescu,
  Karel Dumon, Ivan Duran, Pieter Eendebak, Daniel Egger, Mark Everitt,
  Paco~Mart{\'\i}n Fern{\'a}ndez, Albert Frisch, Andreas Fuhrer, Julien Gacon,
  Gadi, Borja~Godoy Gago, Jay~M. Gambetta, Luis Garcia, Shelly Garion,
  Gawel-Kus, Leron Gil, Juan Gomez-Mosquera, Salvador de~la
  Puente~Gonz{\'a}lez, Donny Greenberg, John~A. Gunnels, Isabel Haide, Ikko
  Hamamura, Vojtech Havlicek, Joe Hellmers, {\L}ukasz Herok, Hiroshi Horii,
  Connor Howington, Wei Hu, Shaohan Hu, Haruki Imai, Takashi Imamichi, Raban
  Iten, Toshinari Itoko, Ali Javadi-Abhari, Jessica, Kiran Johns, Naoki
  Kanazawa, Anton Karazeev, Paul Kassebaum, Vivek Krishnan, Kevin Krsulich,
  Gawel Kus, Ryan LaRose, Rapha{\"e}l Lambert, Joe Latone, Scott Lawrence, Peng
  Liu, Panagiotis Barkoutsos~ZRL Mac, Yunho Maeng, Aleksei Malyshev, Jakub
  Marecek, Manoel Marques, Dolph Mathews, Atsushi Matsuo, Douglas~T. McClure,
  Cameron McGarry, David McKay, Srujan Meesala, Antonio Mezzacapo, Rohit Midha,
  Zlatko Minev, Renier Morales, Prakash Murali, Jan M{\"u}ggenburg, David
  Nadlinger, Giacomo Nannicini, Paul Nation, Yehuda Naveh, Nick-Singstock,
  Pradeep Niroula, Hassi Norlen, Lee~James O'Riordan, Pauline Ollitrault,
  Steven Oud, Dan Padilha, Hanhee Paik, Simone Perriello, Anna Phan, Marco
  Pistoia, Alejandro Pozas-iKerstjens, Viktor Prutyanov, Jes{\'u}s P{\'e}rez,
  Quintiii, Rudy Raymond, Rafael Mart{\'\i}n-Cuevas Redondo, Max Reuter,
  Diego~M. Rodr{\'\i}guez, Mingi Ryu, Martin Sandberg, Ninad Sathaye, Bruno
  Schmitt, Chris Schnabel, Travis~L. Scholten, Eddie Schoute, Ismael~Faro
  Sertage, Yunong Shi, Adenilton Silva, Yukio Siraichi, Seyon Sivarajah,
  John~A. Smolin, Mathias Soeken, Dominik Steenken, Matt Stypulkoski, Hitomi
  Takahashi, Charles Taylor, Pete Taylour, Soolu Thomas, Mathieu Tillet, Maddy
  Tod, Enrique de~la Torre, Kenso Trabing, Matthew Treinish, TrishaPe, Wes
  Turner, Yotam Vaknin, Carmen~Recio Valcarce, Francois Varchon, Desiree
  Vogt-Lee, Christophe Vuillot, James Weaver, Rafal Wieczorek, Jonathan~A.
  Wildstrom, Robert Wille, Erick Winston, Jack~J. Woehr, Stefan Woerner, Ryan
  Woo, Christopher~J. Wood, Ryan Wood, Stephen Wood, James Wootton, Daniyar
  Yeralin, Jessie Yu, Laura Zdanski, Zoufalc, anedumla, azulehner, bcamorrison,
  drholmie, fanizzamarco, kanejess, klinvill, merav aharoni, ordmoj, tigerjack,
  yang.luh, and yotamvakninibm.
\newblock Qiskit: An open-source framework for quantum computing, 2019.

\bibitem{Pow94:proc}
M.J.D. Powell.
\newblock A direct search optimization method that models the objective and
  constraint functions by linear interpolation.
\newblock In S.~Gomez and J-P. Hennart, editors, {\em Advances in optimization
  and numerical analysis}, pages 51--67. Springer Verlag, 1994.

\end{thebibliography}

\section*{Acknowledgements}
This work was supported by Osakidetza that provided the database. The  study protocol was approved  by  the  Clinical Research  Ethics Committee  of  Euskadi (PI2014074), Spain. Informed consent was not obtained because patient health records were made anonymous and de-identified prior to analysis.

\end{document}